\documentclass[journal,twocolumn]{IEEEtran}   
\usepackage{color}
\usepackage{graphicx}
\usepackage{epstopdf}
\usepackage{amssymb}
\usepackage{amsfonts}
\usepackage{amsmath}
\usepackage{color}
\usepackage{booktabs}
\usepackage{algorithm}
\usepackage{algorithmic}
\usepackage{subeqnarray}
\usepackage{cases}
\usepackage{bm}
\usepackage{subfigure,amsmath,amssymb,cite}
\usepackage{stfloats}

\usepackage{float}
\ifCLASSINFOpdf
\else
\fi
\begin{document}

\title{Three High-rate Beamforming Methods for Active IRS-aided Wireless Network }

\author{Feng Shu,~Jing Liu,~Yeqing Lin,~Yang Liu,~Zhilin Chen,\\
 ~Xuehui Wang,~Rongen Dong and Jiangzhou Wang,\emph{ Fellow, IEEE}

\thanks{This work was supported in part by the National Natural Science Foundation of China (Nos.U22A2002, and 62071234), the Hainan Province Science and Technology Special Fund (ZDKJ2021022), and the Scientific Research Fund Project of Hainan University under Grant KYQD(ZR)-21008.
 \emph{(Corresponding authors: Feng Shu and Xuehui Wang)}.}
\thanks{Feng Shu is with the School of Information and Communication Engineering, Hainan University, Haikou, 570228, China, and also with the School of Electronic and Optical Engineering, Nanjing University of Science and Technology, Nanjing, 210094, China (e-mail: shufeng0101@163.com).}
\thanks{Jing Liu, Yeqing Lin, Zhilin Chen, Xuehui Wang and Rongen Dong are with the School of Information and Communication Engineering, Hainan University, Haikou, 570228, China.}

\thanks{Yang Liu is with the School of Electronic and Optical Engineering, Nanjing University of Science and Technology, 210094, China.}

\thanks{Jiangzhou Wang is with the School of Engineering, University of Kent, Canterbury CT2 7NT, U.K. (e-mail: {j.z.wang}@kent.ac.uk).
}

%
}

\maketitle
\begin{abstract}
Due to its ability of breaking the double-fading effect experienced by passive intelligent reflecting surface (IRS), active IRS is evolving a potential technique for future 6G wireless networks. To fully exploit the amplifying gain achieved by active IRS, two high-rate methods, maximum ratio reflecting (MRR) and selective ratio reflecting (SRR) are presented, which are motivated by maximum ratio combining and selective ratio combining. Moreover, both MRR and SRR are in closed-form expressions. To further improve the rate, a maximum approximate-signal-to-noise ratio (Max-ASNR) is first proposed with an alternately iterative infrastructure between adjusting the norm of beamforming vector and its normalized vector. This may make a substantial rate enhancement over existing equal-gain reflecting (EGR). Simulation results show that the proposed three methods perform much better than existing method EGR in terms of rate. They are in decreasing order of rate performance: Max-ASNR, MRR, SRR, and EGR.
\end{abstract}
\section{Introduction}
Since the end of last century, mobile communication has flourished with each passing day, but now faces some challenging problems like high energy consumption and cost. With the advent of  IRS\cite{Wuqingqing2020,Pancunhua2020,shen2019secrecy}, these problems will be addressed without effort due to its passive property and low cost. Moreover, IRS will play an extremely important role in a wide variety of areas including multi-cell MIMO \cite{Pancunhua2020}, spatial modulation\cite{shu2022beamforming_and}, directional modulation\cite{shu2021DM,shu2016robust}, covertness\cite{zhou2021intelligent}, and relay station (RS) \cite{wang2022beamforming}.

 In \cite{shu2022beamforming_and}, an IRS-assisted secure spatial modulation (SM) was proposed to maximize the average secrecy rate (SR) by jointly optimizing passive beamforming at IRS and transmit power at transmitter. In \cite{shen2019secrecy}, an IRS-aided secure wireless network was established and the corresponding method of maximizing SR was presented to jointly optimize the transmit covariance and phase shift matrix of IRS via an efficient alternating structure. \cite{guan2020intelligent} showed that introducing artificial noise (AN) may make an obvious enhancement in the security performance in the IRS-aided setup.
 In IRS-aided MIMO wireless powered communication networks (WPCN) scenario \cite{shi2022}, the secrecy throughput maximization method was proposed by jointly optimizing DL/UL time allocation factor, energy beamforming matrix, information beamforming matrix and DL/UL IRS phase shifts. After DL/UL time allocation factor was given, the original problem was first transformed into a new equivalent form by MSE based method, and then a high quality sub-optimal solution was obtained by alternate iterative optimization algorithm. The results showed that compared with the traditional scheme, the IRS-assisted scheme can improve the security throughput and reduce the ''double near and far'' effect of WPCN system.
In \cite{wang2022beamforming}, an IRS-assisted DF relay network was considered. In order to optimize the beamforming of the RS and the phase of the IRS, the use of alternating iterative structures, based on space domain projection plus maximum ratio combination (MRC) and based on IRS element selection plus MRC, was proposed, respectively.

In a multi-cell scenario with cell-edge users \cite{Pancunhua2020}, a method of maximizing the weighted sum rate  of all users was proposed through jointly optimizing the active precoding matrices at the BSs and the phase shifts at the IRS under unit modulus  and transmit power constraints to improve the cell-edge user performance. Here, phase problem  was addressed by the majorization-minimization and  complex circle manifold methods.
 In \cite{yang2020performance}, a IRS-assisted UAV network system was proposed to reduce the average bit error rate and improve the coverage probability, average capacity and reliability of the system.
 To make an analysis of the performance loss (PL) caused by IRS with phase quantization error, the closed-form expressions of signal to-noise ratio SNR PL, achievable rate, and bit error rate were successively derived under the line-of-sight channels and Rayleigh channels in \cite{dong2022performance}. In a secure wireless network \cite{dong2021active}, a novel alternating method of jointly optimizing the beamformer at transmitter and reflecting coefficient matrix at IRS was designed to harvest a significant secrecy performance gains over the cases with passive IRS and without IRS design.

Recently, active IRS becomes popular due to its amplifying ability. This will lead to the fact that active IRS has a stronger capability to overturn the double-fading effect in wireless fading channels compared to passive IRS in some specific situations depending on the size of IRS and power budget. In \cite{zhi2022active}, the authors made a fair comparison between active and passive IRSs and found that active IRS exceed passive IRS for small-scale or medium-scale IRS with sufficient power budget. In \cite{xu2021resource}, an active IRS-assisted multi-user system was considered, and a method of minimizing the BS transmit power was proposed to jointly optimizes the reflection matrix at IRS and the beamforming vectors at BS to achieve a green wireless communication. In \cite{Long}, based on an active RIS-aided  SIMO system, the amplitude and phase of IRS were separately designed
by sequential convex approximation (SCA) to get the maximum rate at the user.
In \cite{zhang2021active}, active IRS was shown to perform better than passive IRS in small or medium-scale  scenario. Moreover, a closed-form beamforming expression in single-user situation was proposed to simplify the design of beamforming at IRS by combining phase alignment and equal-gain reflecting (EGR). In \cite{lin2022enhancedrate}, based on active IRS, three effective iterative beamforming algorithms GMRR, Max-SNR-FP and Max-SSNR-RR, were proposed to improve the rate at the user.

 Due to the double-fading effect, different product channels have different channel gains and even a part of channels is in deep fading. Deep fading means that those channel gains have extremely small values. In such a situation, EGR in \cite{zhang2021active} allocated the same amplification factor to all channels. It is unreasonable because EGR will treat those deep-fading channels equally and thus degrade the receive signal. Apparently, the EGR  did  not fully exploit the amplifying factor benefit of active IRS. Thus, in this paper, three enhanced beamforming methods at IRS are proposed to improve the data rate performance achieved by active IRS. The main contributions of this paper are summarized as follows:
\begin{enumerate}
 \item To make a full use of the amplifying capacity, the maximum ratio reflecting (MRR) is proposed to optimize both amplitude and phase of  IRS under the transmit power constraints. To reduce the back-haul communication overhead of sending channel state information from BS to IRS, selection ratio reflecting (SRR) algorithm is introduced to activate the $K$ product channels with best channel quality, which is motivated by selective ratio combining (SRC). In accordance with simulation results, the proposed MRR and SRR harvest about one-bit rate gain over existing EGR in the small-scale scenario. More importantly, they have the same low-complexity as EGR.

  \item To further improve the rate, the optimization problem of maximizing the reflected-signal-to-noise ratio (Max-ASNR) is established with a normalized beamforming vector and the corresponding closed-form solution is derived to be a function of the norm of beamforming vector. Then, the computed  beamforming vector is scaled to reach up to the maximum power budget of IRS. Finally, an alternately iterative process is introduced  between the norm and its normalized vector to enhance the rate. Simulation results confirms that the iterative method may converge rapidly within two or three iterations and make an about 2-bit rate improvement over EGR and performs much better than MRR and SRR.

\end{enumerate}

Notations: In this paper, bold lowercase and uppercase letters represent vectors and matrices, respectively. Signs$(\cdot)^H$, $(\cdot)^{-1}$, $tr(\cdot)$, $(\cdot)^{*}$ and $\| \cdot \|$ denote the conjugate transpose operation, inverse operation, trace operation, conjugate operation and 2-norm operation, respectively. The notation $\mathbf{I}_N$ is the $N \times N$ identity matrix. The sign $ E\{ \cdot \} $ represents the expectation operation, and $\text{diag}(\cdot)$ denotes the diagonal operator.

\section{system model}
\begin{figure}[h]
\centering
\includegraphics[width=0.35\textwidth]{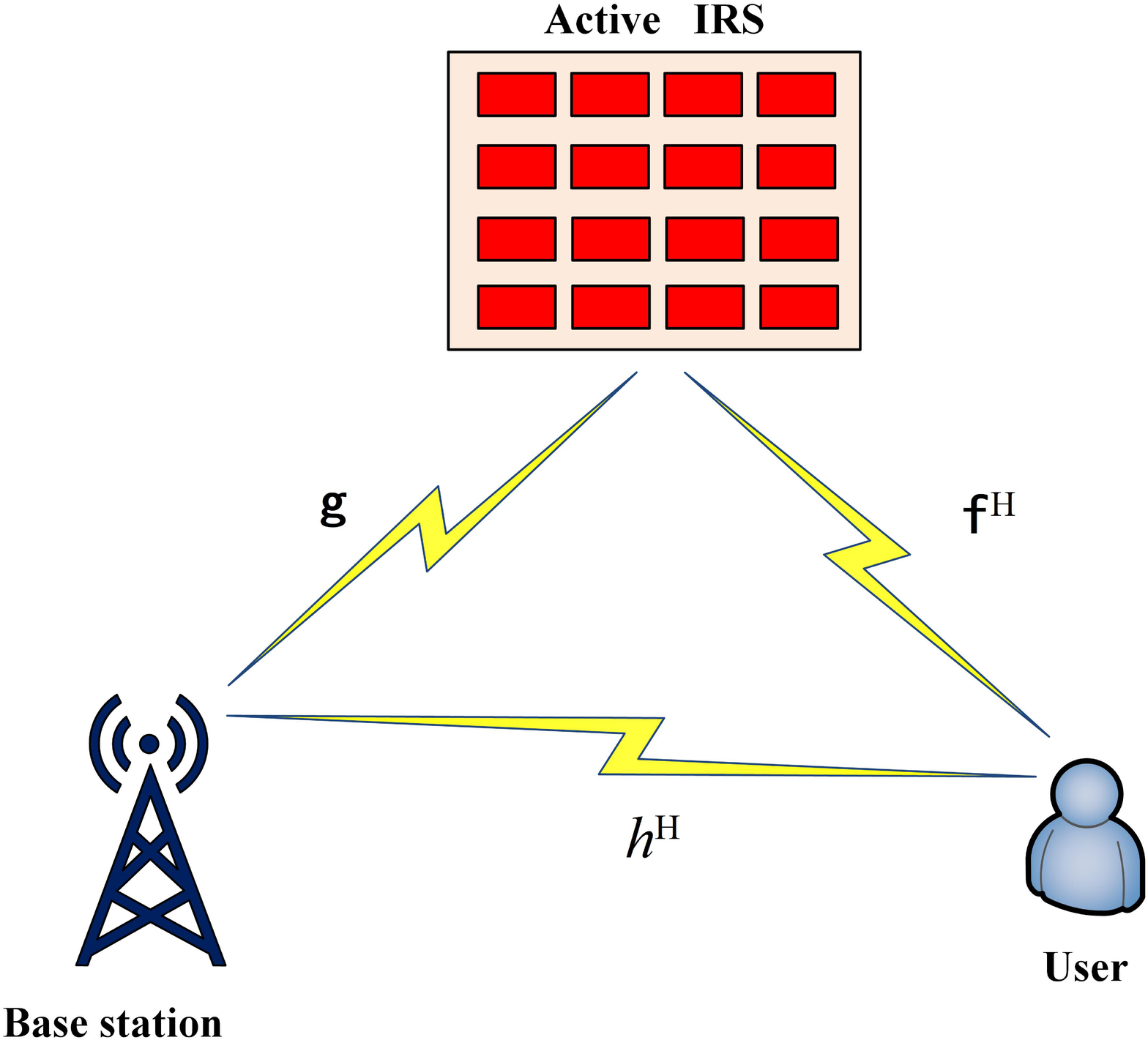}\\
\caption{A typical active IRS-aided wireless network.}\label{active.eps}
\end{figure}
Fig. \ref{active.eps} shows a typical active IRS-aided wireless network. Here, IRS employs $N$ active reflecting elements with element $n$ having amplifying factor $p(n)$. Both base station (BS) and user are equipped with a single antenna. Without loss of generality, there exists a Rayleigh fading among BS, IRS and user. The path-loss exponents of all channels from BS to IRS, IRS to user, and BS to user are larger than two.

The receive signal at IRS is
\begin{align}
\mathbf{x}_I= \mathbf{g}s+\mathbf{w}_I
\end{align}
where $\mathbf{g} \in \mathbb{C}^{N \times 1}$ denotes the  complex channel-gain channel vector from BS to IRS, $s \in \mathbb{C}^{1 \times 1}$ is the data symbol satisfying $E[\|s\|^2] = P_S$, and $\mathbf{w}_I \in \mathbb{C}^{N \times 1}$ is the additive white Gaussian noise (AWGN) with distribution $\mathbf{w}_I \sim \mathcal{C N}\left(0,\sigma_{I}^{2} \right)$.

The signal reflected by IRS is
\begin{align}
\mathbf{y}_I&= \mathbf{P}\mathbf{x}_I
=\mathbf{P}  \mathbf{g}s+ \mathbf{P}  \mathbf{w}_I
\end{align}
where $\mathbf{P} \in \mathbb{C}^{N \times N}$ denotes the diagonal matrix of the active IRS with its diagonal entry $P(n, n)$  being the complex amplifying factor of elements $n$ of IRS. The signal received at user  can be modeled as

\begin{align}
r_U
&= h^Hs+\underbrace{\mathbf{f}^H\mathbf{G}\mathbf{p} s}_{r_I}+ \underbrace{\mathbf{f}^H\mathbf{P} \mathbf{w}_I+w_u}_{w_T}
\end{align}
where $\mathbf{f}^H\in \mathbb{C}^{1 \times N}$ and $ h^H \in \mathbb{C}^{1 \times 1}$ denote the channels from IRS to user, and from BS to user, respectively; $ \mathbf{p}\in \mathbb{C}^{N \times 1}$ denotes column vector form of $\mathbf{P}$, $\mathbf{G}\in\mathbb{C}^{N \times N}$ denotes diagonal matrix form of $ \mathbf{g}\in \mathbb{C}^{N \times 1}$. $w_u$ is the AWGN with distribution $w_u \sim \mathcal{C N}\left(0,\sigma_u^{2} \right)$.

The average total power reflected by IRS is follows
\begin{align}\label{P_I-expre}
E\left\{\left(y_I\right)^H y_I \right\}&=P_{S}\mathbf{p}^H \mathbf{G}^H\mathbf{G}\mathbf{p}+\sigma_{I}^{2} \mathbf{p}^H\mathbf{p}\nonumber\\
&=P_{S}\sum_{n=1}^{N}|p(n)g(n)|^2+\sigma_{I}^{2}\sum_{n=1}^{N}|p(n)|^2
=P_I
\end{align}
where $P_I$ stands for the maximum reflecting power sum at IRS.

For convenience of deriving in the next section, the beamforming vector $\mathbf{p}$ at IRS is factored as the product of its norm $\lambda$ and its normalized vector $\tilde{\mathbf{p}}$
\begin{align}\label{amp-phase}
\mathbf{p}=\tilde{\mathbf{p}}\lambda
\end{align}
where $\lambda=\|\mathbf{p}\|_2$.
The receive total power at the user is given by
\begin{align}
E\left\{\left(r_U\right)^H r_U \right\}&= P_S\| h^H+\mathbf{f}^H\mathbf{G}\mathbf{p}\|_2^2
+\sigma_I^2\mathbf{p}^H\mathbf{F}\mathbf{F}^H\mathbf{p}+\sigma_u^2
\end{align}
By substituting (\ref{amp-phase}) into (\ref{P_I-expre}), we yield
\begin{align}
&E\left\{\left(y_I\right)^H y_I \right\}= P_{S}\lambda^2\tilde{\mathbf{p}}^H\mathbf{G}^H\mathbf{G}\tilde{\mathbf{p}}+\lambda^2\sigma_I^2\tilde{\mathbf{p}}^H\tilde{\mathbf{p}}\nonumber\\
&= P_{S}\lambda^2\sum_{n=1}^{N}|\tilde{p}(n)g(n)|^2+\lambda^2\sigma_{I}^{2}\sum_{n=1}^{N}|\tilde{p}(n)|^2
\end{align}
then we get
\begin{align}\label{lambda}
\lambda&=\sqrt{\frac{P_I}{P_{S}\tilde{\mathbf{p}}^H\mathbf{G}^H\mathbf{G}\tilde{\mathbf{p}}+\sigma_I^2\tilde{\mathbf{p}}^H\tilde{\mathbf{p}}}}
\end{align}
The SNR at the user can be formulated as follows
\begin{align}\label{SNR2}
\text {SNR} &=
 P_{S}\frac{\| h^H+\mathbf{f}^H\mathbf{G}\mathbf{p}\|_2^2} {\sigma_I^2\mathbf{p}^H\mathbf{F}\mathbf{F}^H\mathbf{p}+\sigma_u^2}
\end{align}
The achievable rate at the user can be formulated as follows
\begin{align}
\text {R} &=\log_2(1+\text{SNR})
=\log_2(1+
 P_{S}\frac{\| h^H+\mathbf{f}^H\mathbf{G}\mathbf{p}\|_2^2} {\sigma_I^2\mathbf{p}^H\mathbf{F}\mathbf{F}^H\mathbf{p}+\sigma_u^2})
\end{align}
\section{Proposed MRR, SRR and Max-ASNR}
In what follows, to obtain a higher rate  than conventional method EGR in \cite{zhang2021active}, three high-rate beamformers at IRS, called MRR, SRR and Max-ASNR, are constructed. The third method Max-ASNR is of the iterative structure where the former two methods are with closed-form expressions. Their rate performance exceeds that of EGR.
\subsection{Proposed MRR and SRR}
First, inspired by MRC at receiver, we directly give the following beamforming form
\begin{align}\label{MRR}
\mathbf{p}_{MRR}=\lambda_{MRR}\tilde{\mathbf{p}}=\lambda_{MRR}\frac{\mathbf{G}^H\mathbf{f}}{\|\mathbf{f}^H\mathbf{G}\|_2}\cdot\exp(-j\phi_h)
\end{align}
which is called MRR and $\phi_h$ represents the phase of the direction channel complex gain $h$ from BS to user. Seeing the right side of the above equation, the first term makes all-reflected-path phases alignment and the second one rotates a phase $\phi_h$ to align with the phase of direct path. Substituting the $\tilde{\mathbf{p}}_{MRR}$ into (\ref{lambda}) yields
\begin{align}
\lambda_{MRR}&=\sqrt{\frac{P_I\|\mathbf{f}^H \mathbf{G}\|_2^2}{\left( P_{S}\mathbf{f}^H \mathbf{G}\mathbf{G}^H \mathbf{G}\mathbf{G}^H\mathbf{f}+ \sigma_{I}^{2}\|\mathbf{G}^H \mathbf{f}\|_2^2\right)}}
\end{align}

Now, similar to the SRC in space time coding, we propose a more simple method, called SRR scheme as follows. First, we compute all absolute values of product channel $\mathbf{g}^*(n)\mathbf{f}(n)$, these values are arranged in descending order $\pi$ by sorting methods like the bubble sort algorithm as follows
\begin{align}
\|g(\pi_{1})^*f(\pi_{1})\|_2\ge\cdots\ge\|g(\pi_{N})^*f(\pi_{N})\|_2
\end{align}
from which the  $K$ channels with  largest-gains are chosen to reflect and the remaining $N-K$ ones keep passive or closed. In the above expression, $\pi_n \in~S_N=\{1,~2,\cdots,N\}$ and $\pi$ is a permutation of $S_N$. For convenience of deriving in the following, let us define the new channel vectors as follows
\begin{align}
&\bar{\mathbf{g}}_K=\left[g(\pi_{1}),~\cdots,~g(\pi_{K})\right]^T,
&\bar{\mathbf{f}}_K=\left[f(\pi_{1}),~\cdots,~f(\pi_{K})\right]^T
\end{align}
then, we have the SRR form
\begin{align}\label{MRR}
\mathbf{p}_{SRR}=\lambda_{SRR}\tilde{\mathbf{p}}=\lambda_{SRR}\frac{\bar{\mathbf{G}}_K^H\bar{\mathbf{f}}_K}{\|\bar{\mathbf{f}}_K^H\bar{\mathbf{G}}_K\|_2}\cdot\exp(-j\phi_h)
\end{align}
where $\bar{\mathbf{G}}_K = \text{diag}(\bar{\mathbf{g}}_K)$, Substituting the $\tilde{\mathbf{p}}_{SRR}$ into (\ref{lambda}) yields
\begin{align}
\lambda_{SRR}&=\sqrt{\frac{P_I\|\bar{\mathbf{f}}_K^H \bar{\mathbf{G}}_K\|_2^2}{\left( P_{S}\bar{\mathbf{f}}_K^H \bar{\mathbf{G}}_K \bar{\mathbf{G}}_K \bar{\mathbf{G}}_K^H \bar{\mathbf{G}}_K^H \bar{\mathbf{f}}_K + \sigma_{I}^{2}\|\bar{\mathbf{G}}_K^H \bar{\mathbf{f}}_K\|_2^2\right)}}
\end{align}

Finally, according to our norm-phase framework in (\ref{amp-phase}) and (\ref{lambda}),  the normalized beamforming vector and its norm of EGR in \cite{zhang2021active} can be formulated as follows
\begin{align}
\tilde{\mathbf{p}}_{EGR}=\frac{\mathbf{p}}{\lambda_{EGR}}=\frac{1}{\sqrt{N}}\left[\exp(j\theta_1),~\cdots,~\exp(j\theta_N)\right]^T
\end{align}
where $\theta_n=\arg{(f_ng_n^*)}$, substituting the above expression into (\ref{lambda}) yields
\begin{align}
\lambda_{EGR}=\sqrt{\frac{P_I N}{P_S\sum_{n=1}^{N}|g(n)|^2+N\sigma_I^2}}
\end{align}
\subsection{Proposed Max-ASNR }
The total average receive  SNR at user is written as
\begin{align}\label{SNR-1}
&\mathbf{SNR}(\mathbf{p},\lambda)=P_S \bullet \nonumber\\ &\frac{\mathbf{p}^H\mathbf{G}^H\mathbf{f}\mathbf{f}^H\mathbf{G}\mathbf{p}+h^H\mathbf{p}^H\mathbf{G}^H\mathbf{f}+h\mathbf{f}^H\mathbf{G}\mathbf{p}+h^Hh}{\sigma_I^2 \mathbf{p}^H \mathbf{F} \mathbf{F}^H \mathbf{p}+\sigma_u^2}
\end{align}

which may be rewritten as
\begin{align}\label{ASNR-1}
&\mathbf{SNR}(\mathbf{p},\lambda)=P_S \bullet \nonumber\\ &\frac{\mathbf{p}^H\mathbf{G}^H\mathbf{f}\mathbf{f}^H\mathbf{G}\mathbf{p}+h^H\mathbf{p}^H\mathbf{G}^H\mathbf{f}+h\mathbf{f}^H\mathbf{G}\mathbf{p}+h^Hh}{ \mathbf{p}^H( \sigma_I^2 \mathbf{F} \mathbf{F}^H +\lambda^{-2}\sigma_u^2\mathbf{I}_{N})\mathbf{p}}
\end{align}

Let us define $\mathbf{D}=\sigma_I^2 \mathbf{F} \mathbf{F}^H +\lambda^{-2}\sigma_u^2\mathbf{I}_{N}$,
with $\mathbf{D}^{\frac{1}{2}}(n)=\sqrt{\sigma_I^2|f(n)|^2+\lambda^{-2}\sigma_u^2}$, then
\begin{align}\label{ASNR-1}
&\mathbf{SNR}(\mathbf{p},\lambda)=P_S \bullet \nonumber\\ &\frac{\mathbf{p}^H\mathbf{G}^H\mathbf{f}\mathbf{f}^H\mathbf{G}\mathbf{p}+h^H\mathbf{p}^H\mathbf{G}^H\mathbf{f}+h\mathbf{f}^H\mathbf{G}\mathbf{p}+h^Hh}{ \mathbf{p}^H\mathbf{D}\mathbf{p}}
\end{align}

with  $\mathbf{p}'=\mathbf{D}^{\frac{1}{2}}\mathbf{p}$ with  ${\mathbf{D}^{\frac{1}{2}}}^H\mathbf{D}^{\frac{1}{2}}=\mathbf{D}$. Substituting them in (\ref{ASNR-1}) yields
\begin{align}\label{SNR-1}
\mathbf{SNR}(\mathbf{p}',\lambda)=P_S \frac{\mathbf{p'}^H{\mathbf{D}^{-\frac{1}{2}}}^H\mathbf{G}^H\mathbf{f}\mathbf{f}^H\mathbf{G}\mathbf{D}^{-\frac{1}{2}}\mathbf{p}'+\mathbf{A}+h^Hh}{ \mathbf{p}'^H\mathbf{p}'}
\end{align}

where $\mathbf{A}=h^H\mathbf{p}'^H{\mathbf{D}^{-\frac{1}{2}}}^H\mathbf{G}^H\mathbf{f}+h\mathbf{f}^H\mathbf{G}\mathbf{D}^{-\frac{1}{2}}\mathbf{p}'$
.

In order to simplify the above objective function, the term $h^Hh$  of  the numerator on right side is omitted, called approximate SNR (ASNR) without term $h^Hh$,  and we have
\begin{align}\label{SNR-1}
\mathbf{ASNR}(\mathbf{p}',\lambda)=P_S \frac{\mathbf{p'}^H{\mathbf{D}^{-\frac{1}{2}}}^H\mathbf{G}^H\mathbf{f}\mathbf{f}^H\mathbf{G}\mathbf{D}^{-\frac{1}{2}}\mathbf{p}'+\mathbf{A}}{ \mathbf{p}'^H\mathbf{p}'}
\end{align}

Maximizing the above function $\mathbf{ASNR}(\mathbf{p}',\lambda)$ is cast as
\begin{align}
&\max_{\mathbf{p}',\lambda} ~~\mathbf{ASNR}(\mathbf{p}',\lambda)\\
~&\text{s.t.} ~~~{\mathbf{p}'}^H {\mathbf{p}'}=b,\nonumber
\end{align}
where $b$ is a constant, let us define matrix
\begin{align}
\mathbf{C}={\mathbf{D}^{-\frac{1}{2}}}^H\mathbf{G}^H\mathbf{f}\mathbf{f}^H\mathbf{G}\mathbf{D}^{-\frac{1}{2}}
\end{align}
then (24) reduces to
\begin{align}
&\max_{\mathbf{p}',\lambda} ~~\widetilde{\mathbf{{ASNR}}}(\mathbf{p}',\lambda)=\mathbf{p'}^H\mathbf{C}\mathbf{p}'+\mathbf{A}\\
~&\text{s.t.} ~~~{\mathbf{p}'}^H {\mathbf{p}'}=b,\nonumber
\end{align}

Removing the unit-mod constraint forms the following unconstrained optimization problem
\begin{align}
&\max_{\mathbf{p}',\lambda} ~~\mathbf{p'}^H\mathbf{C}\mathbf{p}'+\mathbf{A}
\end{align}

Given the value of $\lambda$, taking the derivative of objective function $\widetilde{\mathbf{{ASNR}}}(\mathbf{p}',\lambda)$ with respect to  $\mathbf{p}'$ and setting it equal zero generate the following equation
%

\begin{align}
\frac{\partial \widetilde{\mathbf{{ASNR}}}(\mathbf{p}',\lambda)}{\partial \mathbf{p'}^* }=\mathbf{C}\mathbf{p}'+h^H {\mathbf{D}^{-\frac{1}{2}}}^H \mathbf{G}^H\mathbf{f}=0
\end{align}
which  gives the solution to $\mathbf{p}'$ as follows

\begin{align}
\mathbf{p}'_{ASNR}(\lambda)=\frac{-h^H\mathbf{C}^{-1}{\mathbf{D}^{-\frac{1}{2}}}^H\mathbf{G}^H\mathbf{f}}{\|-h^H\mathbf{C}^{-1}{\mathbf{D}^{-\frac{1}{2}}}^H\mathbf{G}^H\mathbf{f}\|_2}
\end{align}
then we get $\mathbf{p}_{ASNR}(\lambda)$
\begin{align}\label{30}
\mathbf{p}_{ASNR}(\lambda)&=\mathbf{D}^{-\frac{1}{2}}\mathbf{p}'_{ASNR}(\lambda)
=\frac{-h^H\mathbf{D}^{-\frac{1}{2}}\mathbf{C}^{-1}{\mathbf{D}^{-\frac{1}{2}}}^H\mathbf{G}^H\mathbf{f}}{\|-h^H\mathbf{C}^{-1}{\mathbf{D}^{-\frac{1}{2}}}^H\mathbf{G}^H\mathbf{f}\|_2}
\end{align}
with its normalized solution
\begin{align}\label{Phase-ASNR}
\tilde{\mathbf{p}}_{ASNR}(\lambda)=\mathbf{p}_{ASNR}(\lambda)\|\mathbf{p}_{ASNR}(\lambda)\|_2^{-1}
\end{align}

Substituting the above normalized in (\ref{lambda}),  we get the $\lambda_{Max-ASNR}$.
\begin{align}\label{lambda-ASNR}
&\lambda_{Max-ASNR}(\widetilde{\mathbf{p}}_{ASNR})\nonumber\\&=\sqrt{\frac{P_I}{P_{S}\tilde{\mathbf{p}}_{ASNR}^H\mathbf{G}^H\mathbf{G}\tilde{\mathbf{p}}_{ASNR}+\sigma_I^2\tilde{\mathbf{p}}_{ASNR}^H\tilde{\mathbf{p}}_{ASNR}}}\nonumber\\
&=\sqrt{\frac{P_I}{P_{S}\sum_{n=1}^{N}|\tilde{p}_{ASNR}(n)g(n)|^2+\sigma_{I}^{2}\sum_{n=1}^{N}|\tilde{p}_{ASNR}(n)|^2}}
\end{align}
which is called the Max-ASNR method. This method actually maximizes the ASNR  at user.
 Then an alternate iterative process between  (\ref{Phase-ASNR}) and (\ref{lambda-ASNR}) is performed until convergence. Its detailed iterative process is listed in Algorithm I.

\begin{table}[htbp]\normalsize
 \centering
 \label{tab:pagenum}
 \begin{tabular}{p{240pt}}
  \toprule
   \textbf{Algorithm I} Proposed Max-ASNR method\\
  \midrule
  Step 1: Place the initial value $\lambda_{MRR}$ into (\ref{30}), then we get the value of $\tilde{\mathbf{p}}_{ASNR}$ by (\ref{Phase-ASNR}).\\
  Step 2: Substitute $\tilde{\mathbf{p}}_{ASNR}$ into (\ref{lambda-ASNR}),
  and output the new value of $\lambda_{Max-ASNR}$. \\
  Step 3: \textbf{Repeat} Steps 1-2\\
  \textbf{Until} $\lambda_{new}-\lambda_{old} \leq \epsilon$\\
  \bottomrule
 \end{tabular}
\end{table}

\subsection{Complexity analysis}
In what follows, we will make a detailed analysis concerning the complexities of the three proposed methods with EGR,  AO in [17], three methods in [18] and SCA in [16]  as the performance analysis. There is an increasing order in complexity (FLOPs): EGR ($\mathcal{O}(N)$ FLOPs) $\approx$ MRR ($\mathcal{O}(N)$ FLOPs) $\leq$ SRR ($\mathcal{O}(N^2)$ FLOPs) $\leq$ GMMR ($\mathcal{O}(N^3)$ FLOPs) $\approx$ Max-SSNR-RR ($\mathcal{O}(N^3)$ FLOPs) $\approx$ Max-ASNR ($\mathcal{O}(N^3)$ FLOPs) $\leq$ SCA ($\mathcal{O}(N^{3.5})$ FLOPs) $\leq$ AO ($\mathcal{O}(N^{4.5})$ FLOPs) $\approx$ Max-SNR-FP ($\mathcal{O}(N^{4.5})$ FLOPs). In particular, the proposed iterative Max-ASNR achieves far lower-complexity than AO and SCA.

\section{Simulation and Numerical results}
In this section, numerical simulation results are conducted to evaluate the rate and convergent performance of our proposed methods. Simulation parameters are set as follows: $\sigma_I^2$ = $ \sigma_u^2 $ = -70 dBm,~$P_I$ = 30 dBm,~and $P_S$ = 15 dBm. BS, user, and active IRS are located at (0 m, 0 m), (150 m, 0 m)~and (100 m, 30 m). The path loss exponents from BS to IRS, IRS to user, and BS to user are 2.3, 2.3 and 3.8 respectively.

\begin{figure*}[htbp] 
\begin{minipage}[t]{0.34\linewidth} 
\centering
\includegraphics[width=2.5in, height=1.8in]{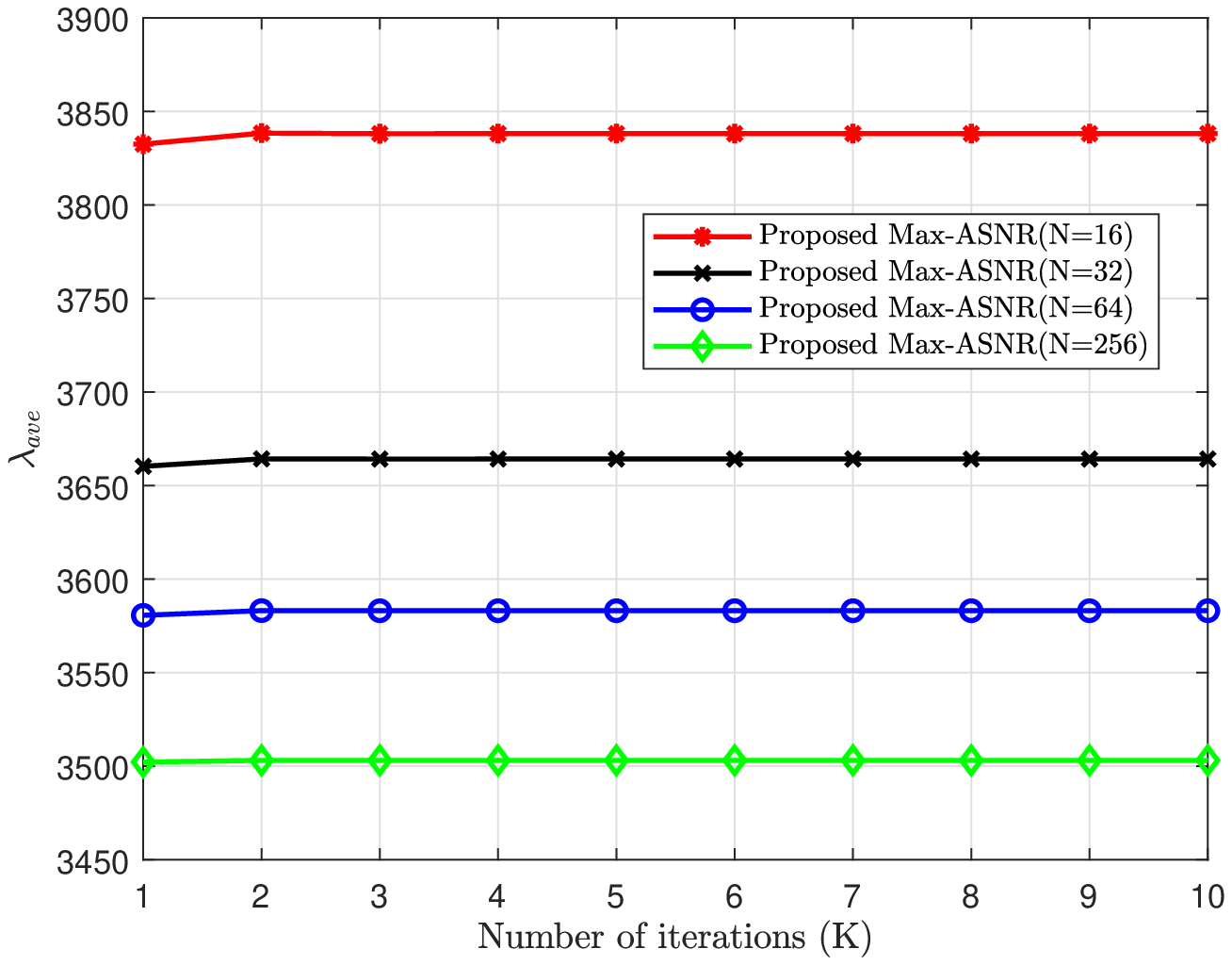} 
\caption{Convergent curves of proposed Max-ASNR method.} 
\label{newdiedai} 
\end{minipage}%
\begin{minipage}[t]{0.34\linewidth}
\centering
\includegraphics[width=2.5in, height=1.8in]{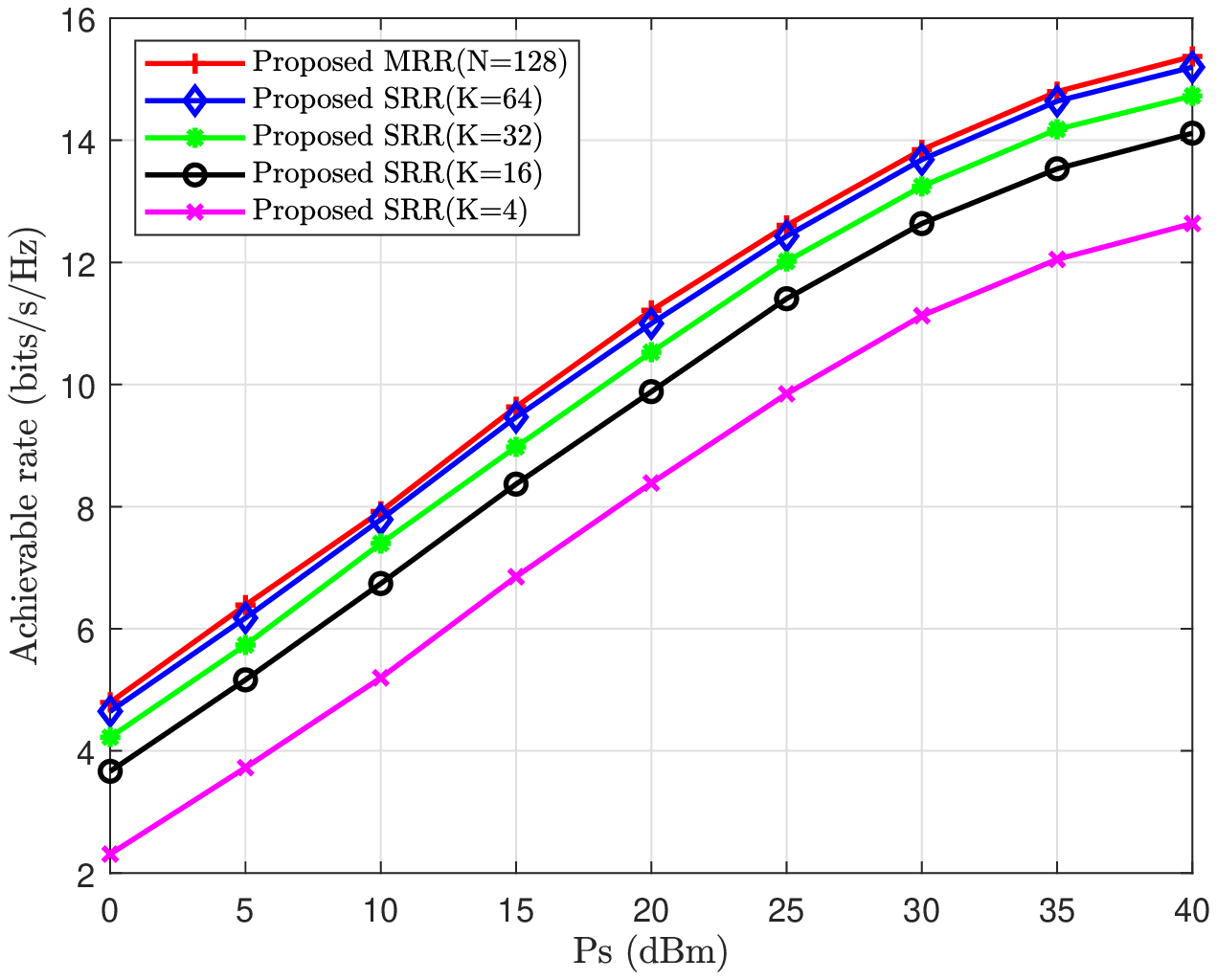}
\caption{Performance of SRR for different $K$.}
\label{SRR}
\end{minipage}%
\begin{minipage}[t]{0.34\linewidth}
\centering
\includegraphics[width=2.5in, height=1.8in]{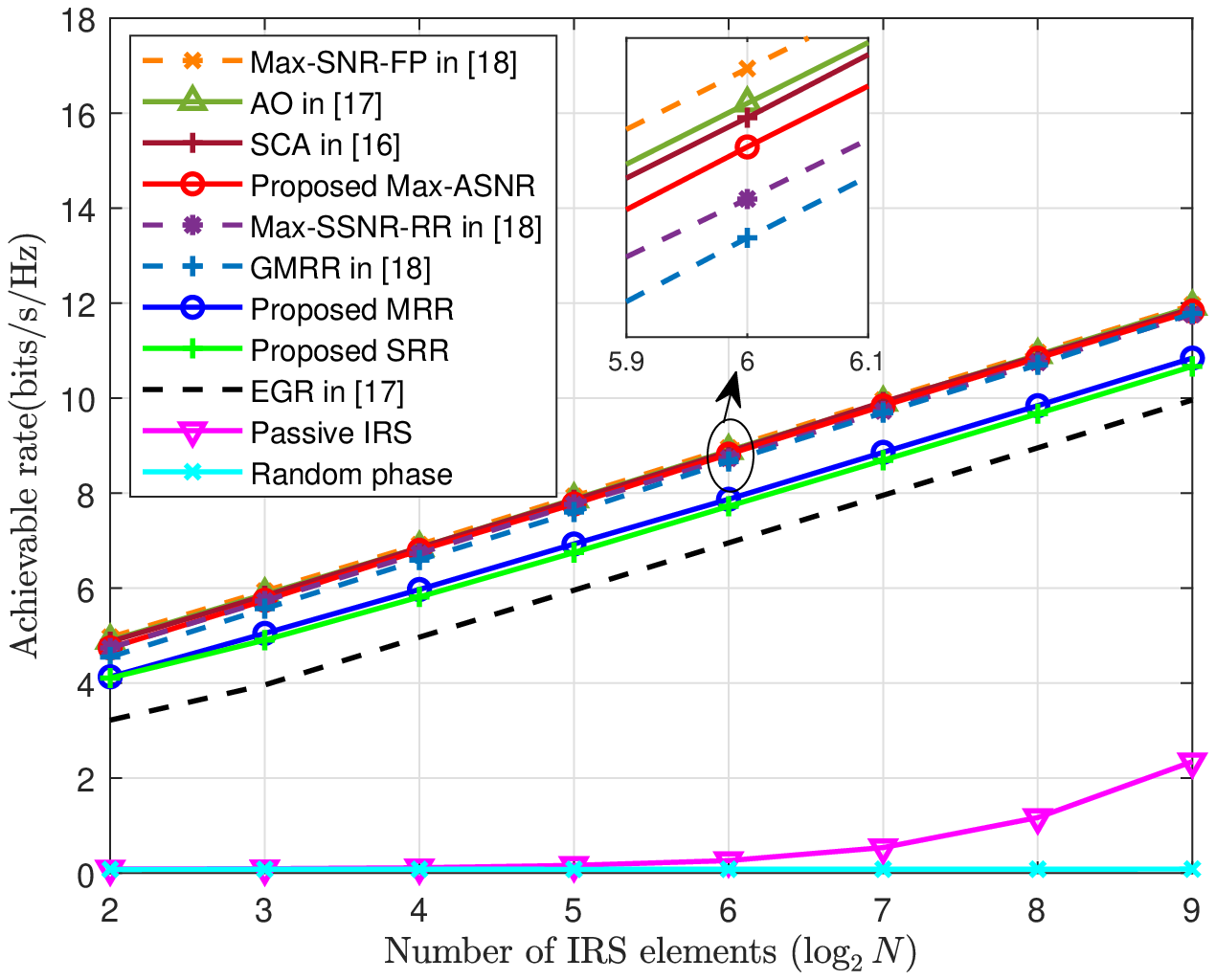}
\caption{Rate performance versus $N$ of the proposed three methods.}
\label{5}
\end{minipage}
\end{figure*}

Fig. 2 shows the convergent curve of the proposed Max-ASNR. From Fig. 2, it is seen that the Max-ASNR, requires  about two or three iterations to converge to the error floor.  Thus, the proposed method has the property of a rapid convergence.


Fig. 3 shows the curves of rate versus the transmit power at BS of the proposed SRR for  different values of $K$. From Fig. 3, it is seen that as  the value of $K$ varies from 4 to 64,  the rate increases gradually. When half of the number of IRS are selected, the rate is very close to that of MRR. Therefore, $K=N/2$ is a feasible choice with  negligible rate loss. In particular,  compared to MRR, this approach can significantly reduce communication overhead from BS to IRS.

Fig. 4 plots the rate versus the number of IRS elements of the proposed three methods with SCA in \cite{Long}, EGR and AO in \cite{zhang2021active}, three methods in \cite {lin2022enhancedrate}, passive IRS and random phase as performance benchmarks. From this figure, it is seen that all methods have a decreasing order in rate as follows: Max-SNR-FP, AO, SCA, Max-ASNR, Max-SSNR-RR, GMMR, MRR, SRR, and EGR. Taking them complexities into account, we can make a summary as follows. The proposed two low-complexity  closed-form methods MRR and SRR achieve about one-bit rate gains over EGR. The former is the same low-complexity as EGR while the latter may  save a substantial communication overhead over EGR with an one-order higher  complexity. The proposed iterative method Max-ASNR makes a good balance among rate performance and computational  complexity compared with existing iterative methods like AO and SCA.

\section{Conclusions}
In this paper, we have made an investigation of beamforming methods of optimizing the amplitude and phase of IRS for active  IRS-assisted three-node networks. To make full use of amplifying property of active IRS to enhance the rate performance, three high-performance  methods Max-ASNR, MRR and SRR were proposed.
The proposed two low-complexity  closed-form methods  MRR and SRR achieve about one-bit rate gain over EGR.  The former is the same low-complexity as EGR while the latter may save a substantial communication  overhead over EGR with an one-order higher complexity. The proposed iterative Max-ASNR makes a good balance between rate and complexity compared to existing iterative methods like AO and SCA.


\ifCLASSOPTIONcaptionsoff
  \newpage
\fi

\bibliographystyle{IEEEtran}
\bibliography{IEEEfull,reference}
\end{document}